# A Novel Gravity–Quasi-Laplacian Approach to Identifying Influential Nodes in Complex Networks

Shima Esfandiari, Seyed Mostafa Fakhrahmad

School of Electrical and Computer Engineering, Shiraz University, Shiraz, Iran

**Abstract:**

Identifying influential nodes in complex networks is a fundamental challenge with broad applications in areas such as social network analysis, communication infrastructure, transportation systems, and information networks. Existing ranking methods typically rely on combinations of structural features—such as degree, k-shell index, and neighborhood connectivity—to estimate a node's importance. However, many of these approaches suffer from key limitations, including insufficient accuracy, low resolution in distinguishing nodes with similar influence, dependence on tunable parameters, and high computational complexity, which restrict their practicality in large-scale or real-world networks. This study introduces a new ranking framework that integrates a quasi-Laplacian structural measure with a gravity-inspired aggregation process. The core idea is to construct a strengthened representation of each node's structural role using only simple yet informative attributes—namely degree and k-shell index—and then evaluate its local influence through a short-range interaction mechanism. The proposed approach is designed to be free of tunable parameters, interpretable, and computationally efficient, requiring only a small fixed gravity radius (R=3), which makes it suitable for large and diverse networks. Experiments conducted on nine real-world networks and compared against eight state-of-the-art methods demonstrate that the proposed framework consistently outperforms existing techniques in terms of accuracy, resolution, and computational simplicity. These results highlight the effectiveness of the gravity–quasi-Laplacian paradigm as a reliable and scalable tool for identifying influential nodes in complex networks.

## 1. Introduction

Advances in complex network science have enabled more accurate analysis and management of real-world systems. Many natural and human phenomena—such as user interactions in social networks, transportation mechanisms, computer infrastructures, disease spread, and even biological and neural processes—can be effectively modeled and examined using network-based representations [1]. Although real networks often exhibit irregular and multilayered structures, the framework of complex networks provides an appropriate tool for simplifying and analyzing their behavior [2-11]. Applications span a wide range of areas, including link prediction, community detection, recommender systems, information flow management, epidemic containment, rumor source tracking, and improving web data quality [2,10].

Among network components, some nodes play a more dominant role in the network's dynamics, known as influential nodes. For example, in social networks, the rapid diffusion of information—whether accurate or false—makes it crucial to identify individuals who can slow or redirect the spread [12]. In epidemiological networks, identifying individuals whose quarantine can stop further transmission is vital [11]. In marketing, determining the users with the highest dissemination potential is essential for maximizing advertising efficiency [13]. Overall, identifying such key nodes is a major part of solving many applied problems.

Existing methods for identifying influential nodes generally fall into two categories: selecting a set of nodes that collectively maximize influence [14], and evaluating the spreading capability of nodes individually to produce a ranking [2-9,18-23]. This study focuses on the latter. Although diffusion-based models provide highly accurate evaluations of spreading power, they are computationally expensive due to their simulation-based nature. Therefore, there is a need for metrics that are both cost-efficient and effective.

Researchers have proposed various methods based primarily on structural graph features. Foundational metrics such as degree [15], K-shell index [16], and betweenness centrality [17] are often combined with other properties to produce better results. Some metrics rely on local network information, while others incorporate global structural characteristics. Combining different features allows methods to achieve more comprehensive perspectives and strong performance across different networks. Such methods should not only achieve high accuracy in identifying influential nodes but also distinguish between nodes with similar spreading capabilities—that is, provide high resolution. Since many real networks are large, computational efficiency is also necessary.

Methods for ranking influential nodes fall into two main groups: (1) interpretable formula-based methods that rely solely on structural features of the graph [18–23], and (2) methods that use neural networks, GCNs, GNNs, or MCDM techniques [2,24]. This work belongs to the first group and focuses on computing interpretable, transparent, and understandable metrics without requiring complex deep-learning models, which reduces computational cost and increases result interpretability. The proposed method is based on a gravity-inspired model and a quasi-Laplacian index that incorporates both local (degree) and global (K-shell) metrics. Comprehensive evaluations show improvements in accuracy, resolution, and computational efficiency compared to eight recent methods on nine real-world networks.

The main innovations of the proposed method can be summarized as follows.

(1) A quasi-Laplacian operator is introduced to construct the HQC index, enabling the method to jointly capture the direct structural contribution of a node and the amplified effect of its local and shell-based connectivity.

(2) Unlike conventional gravity-based approaches that may require a larger or adjustable interaction radius, the proposed method fixes the gravity radius at $R=3$, which reduces computational cost while preserving effective neighborhood information relevant for influence spreading.

(3) The proposed method is free of tunable parameters and requires no user-defined input other than the fixed radius $R=3$, making it easy to apply across different networks without additional tuning.

The remainder of the paper is organized as follows: Section 2 reviews previous work and the comparative methods; Section 3 introduces the proposed method and its flowchart; Section 4 presents the datasets, explains the SIR model [25] (used to assess true spreading capability), and describes the evaluation criteria; Section 5 reports the comparative results from multiple perspectives; and Section 6 concludes the work and outlines directions for future research.

## 2. Related work

This section introduces the two fundamental structural features—degree and K-shell—and describes recent interpretable methods that rely on these features. All methods presented here are included in the comparative evaluation in Section 5.

### 2-1- Degree Centrality (DC)

The degree centrality is the simplest influence metric and relies solely on local information. Its value equals the number of direct neighbors of a node.

### 2-2- K-Shell (KS)

The K-shell index is a layer-by-layer decomposition method based on degrees. All nodes with degree one are first removed and assigned ks-index = 1. This iterative process continues until no degree-one nodes remain. Then the index increases by one, nodes with degree two are removed, and the procedure continues until all nodes receive an index.

### 2-3- Global Structure Model (GSM)

This method [18] defines self-influence and global influence for each node. The self-influence is derived from the KS value, while the global influence is computed using the sum of KS values of neighbors weighted by inverse distances:

$$GSM_i = SI_i \times GI_i = e^{\frac{ks_i}{n}} \times \sum_{i \neq j} \frac{\mathrm{ks}_j}{d_{ij}} \quad (1)$$

**2-4- Hybrid Global Structure Model (HGSM)**

Similar in structure to GSM, but with two modifications [19]: (1) the self-influence component incorporates both degree and KS, and (2) the global component includes statistical properties of the network such as average degree and average KS:

$$HGSM_i = SI_i \times GI_i = e^{\frac{ks_i * \mathrm{k}_i}{n}} \times \sum_{i \neq j} \frac{e^{\frac{\mathrm{ks}_j * \mathrm{k}_j}{n}}}{d_{ij}^{\left\lceil \log_2 < e^{\frac{\mathrm{ks} * k}{n}} > \right\rceil}} \quad (2)$$

Evaluating all nodes in the GI component results in relatively high computational complexity.

**2-5- Local and Global Index (LGI )**

This metric [20] incorporates both local and global components. The local part measures connectivity among neighbors and involves tunable coefficients. The global part is based on the sum of KS values of a node and its neighbors.

$$R(i,j) = \frac{|N(j) \cap N(i)|}{k_i}$$

$$LC(i) = \sum_{j \in N_i} R(i,j)$$

$$LI(i) = \alpha \frac{k_i}{\sum_{j \in N_i} k_j} + \beta \frac{LC(i)}{\sum_{j \in N_i} LC(j)}$$

$$LGI(i) = \sum_{j \in N_i} LI(j) * GI(j) \quad (3)$$

**2-6- GGC [21]**

This gravity-inspired method combines degree and clustering coefficient as the mass in the gravity equation. The distance is taken as shortest-path length, and evaluation is limited to a radius equal to half of the network diameter:

$$\mathrm{GGC(i)} = \sum_{d_{ij} \le R,\ \ j \ne i} \frac{e^{-\alpha\left(LC_i + LC_j\right)} k_i k_j}{d_{ij}^2} \tag{4}$$

**2-7- DNC [22]**

This metric combines node degree with the sum of clustering coefficients of neighbors, controlled by a tunable parameter α:

$$slc_i = \sum_{j \in \mathrm{N}_i} LC_j$$

$$DNC(i) = k_i + \alpha \sum_{j \in \mathrm{N}_i} LC_j\ , \alpha > 0 \tag{5}$$

**2-8- NPIC [23]**

This method uses degree and KS and accounts for the effect of all nodes on each node. Two tunable parameters α and β (between 0.1 and 1) are included:

$$NPIC(i) = \frac{(ks_i * k_i) + \alpha}{n} * \sum_{i \ne j} \frac{(ks_j * k_j) + \beta}{d_{ij}} \tag{6}$$

Although these methods offer valuable insights, they have limitations. For example, HGSM, GSM, and NPIC have quadratic time complexity, while DNC, GGC, and LGI require parameter tuning. Therefore, a method with low time complexity, strong interpretability, and a fixed, small interaction radius is needed. The present work addresses these issues by proposing the QGC method with a fixed radius of R=3, achieving high accuracy and resolution on a broad range of benchmark networks.

Although the above methods provide valuable perspectives for identifying influential nodes, they differ in terms of structural information usage, computational cost, ranking resolution, and parameter dependence. Classical centrality measures are generally simple and interpretable, but they often fail to capture higher-order structural effects related to spreading dynamics. More recent hybrid and gravity-based methods usually achieve better ranking performance by incorporating neighborhood interactions and distance information; however, some of them suffer from high computational complexity, while others require tunable parameters or network-specific settings. To make this comparison clearer, a summary of representative methods is provided in Table 1. As shown, there is still a need for a method that combines high accuracy, strong discriminative power, low computational complexity, and fixed non-tunable settings. The proposed QGC method is designed to address this gap by employing a fixed radius of R=3 and a simple yet effective gravity-based formulation.

Table 1- Comparison of representative influential-node ranking methods

| Method | Structural basis | Main strength | Main limitation | Complexity | Parameter tuning |
|---|---|---|---|---|---|
| **DC** | Degree of node | Very simple and fast | Considers only immediate neighbors | $O(n)$ | No |
| **BC** | Shortest-path bridging role | Captures global influence pathways | High computational cost | $O(nm)$ | No |
| **CC** | Global distance to other nodes | Reflects overall reachability | Sensitive to disconnected structures | $O(nm)$ | No |
| **K-shell** | Core-periphery decomposition | Effective hierarchical representation | Limited resolution for nodes in same shell | $O(m)$ | No |
| **GGC** | Gravity-based interaction | Good accuracy and interpretability | Depends on parameter settings | $O(nlogn)$ | Yes |
| **DNC** | Neighborhood-based structural ranking | Improves local influence estimation | Requires parameter adjustment | $O(n < k >^2)$ | Yes |
| **LGI** | Local gravity interaction | Combines distance and local structure | Radius or weights may need tuning | $O(m < \mathrm{k} >)$ | Yes |
| **HGSM** | Hybrid gravity-shell strategy | Strong ranking performance | Relatively high computational cost | $O(n^2)$ | No |
| **GSM** | Gravity-based spreading measure | Good empirical effectiveness | Computationally expensive | $O(n^2)$ | No |
| **NPIC** | Improved neighborhood/gravity centrality | High ranking capability | Relatively high complexity | $O(n^2)$ | No |
| **QGC** | HQC-based gravity with fixed radius R=3 | Accurate, high-resolution, interpretable, low tuning burden | Fixed radius may not be optimal for every network | $O(n < k >^3)$ | No |

## 3. Proposed method

In the proposed method, influential nodes are identified based on a combination of local structural properties, shell position, and Newton's gravitational law. The central idea is to first define a quasi-Laplacian metric for each node—a metric capable of capturing both the local intensity and the structural role of the node within its immediate neighborhood. To this end, the HQC index is introduced (Equation 1). By combining the degree and K-shell index of each node, along with the aggregated values of these features among its direct neighbors, HQC provides an enhanced representation of the node's structural strength.

The structure of HQC consists of three components: a quadratic term, a linear term, and a neighborhood-dependent term. The quadratic component emphasizes the central power of the node; the linear component maintains intermediate-scale distinctions and prevents the loss of resolution among mid-level nodes; and the neighborhood component strengthens the contribution of the node's local environment to its spreading potential. Together, these components operate similarly to a quasi-Laplacian operator on the graph, meaning that the value of each node and its surrounding values are treated as a local field and structurally combined.

$$HQC(i) = (k_i + ks_i)^2 + (k_i + ks_i) + 2\sum_{j \in Ni}(k_j + ks_j) \tag{7}$$

After HQC is computed for all nodes, it is used as the "network mass" in a gravity-based model. At this stage, the QGC index is defined (Equation 8), which quantifies the influence of each node on surrounding nodes within a fixed radius of R=3. This choice of radius is motivated by the sensitivity analysis presented in Section 5-5, which shows that R=3 provides the best average performance across the majority of tested networks while still preserving locality and low computational cost. Limiting the interaction radius to this fixed local range keeps the computation lightweight and captures the most relevant neighborhood influence information, which is particularly important for the early-stage structural effects reflected by diffusion processes such as the SIR model.

In QGC, the impact of nodes within radius of three is calculated based on the product of their masses (HQC values) and the inverse square of their shortest-path distances. Therefore, nodes that themselves have high structural strength and also lie in the vicinity of other strong nodes will receive a higher QGC score. This mechanism captures the idea that a node's influence arises not only from its intrinsic properties but also strongly from the structural environment surrounding it.

$$QGC(\mathrm{i}) = \sum_{d_{ij} \le R,\ j \ne i} \frac{HQC(i)HQC(j)}{{d_{ij}}^2}\ , supposed\ R = 3 \tag{8}$$

Finally, the QGC value of each node is taken as its influence index. By accurately combining local power, shell depth, and short-range structural interactions, the method provides a realistic estimation of a node's spreading capability within the network. Nodes that score highly under this index typically lie in structurally critical regions and are capable of initiating or steering early diffusion waves—whether related to epidemics, rumors, or information flow—with greater intensity. Fig. 1 illustrates the overall procedure for computing node spreading power using the QGC method.

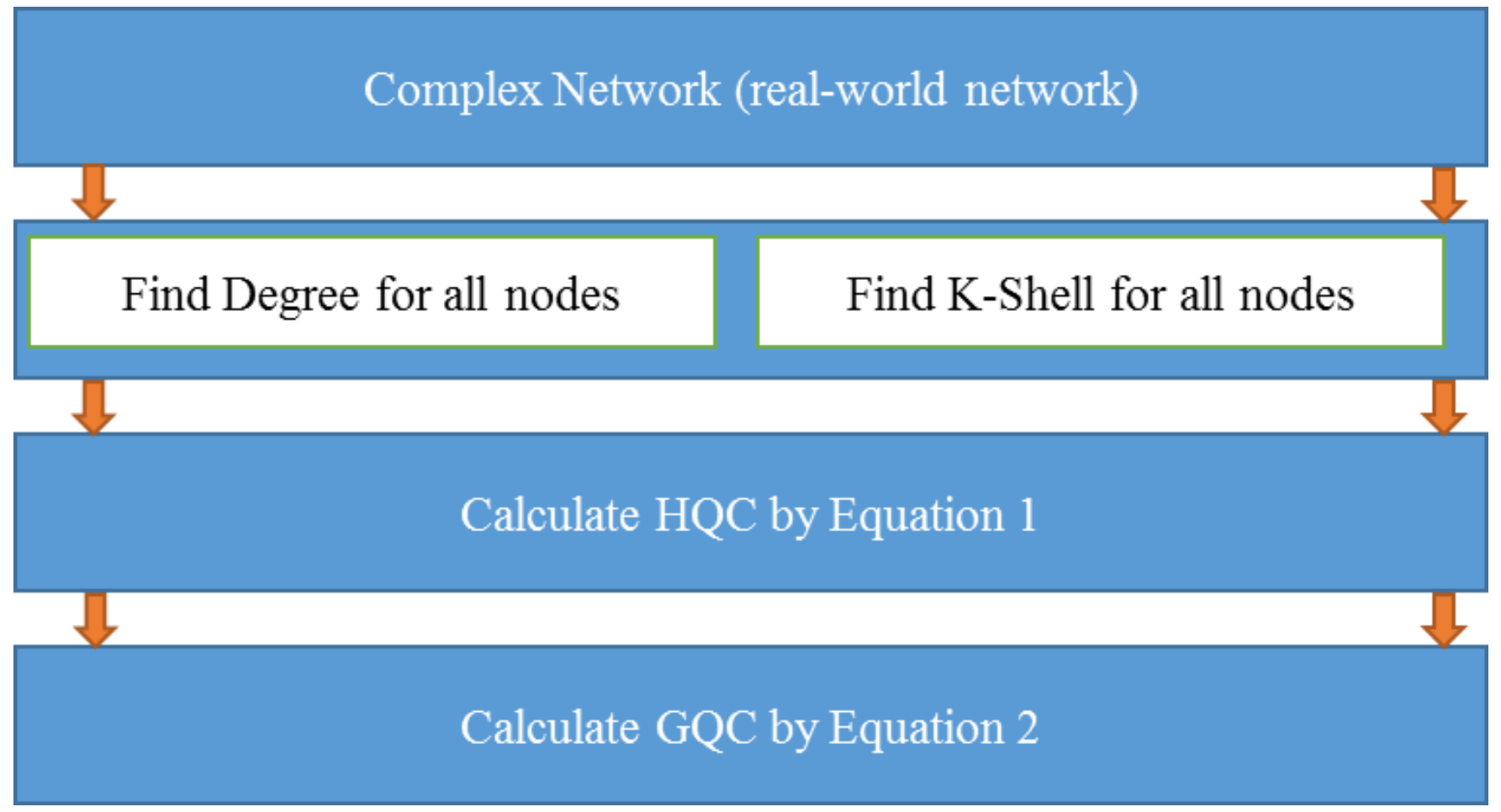

Fig 1. Flow chart of the proposed method

## 4. Dataset and evaluation metrics

### 4-1- SIR model

To assess the spreading capability of nodes, the classical SIR model is employed. In this model, every node can be in one of three states—susceptible (S), infected (I), or recovered (R)—and transitions occur probabilistically based on infection probability β and recovery probability α. For each experiment, the process begins by infecting a single node while all others are considered susceptible. The infection then propagates through the network as infected nodes transmit the disease to their susceptible neighbors with probability β, and subsequently recover with probability α. The relative number of recovered nodes at the end of the simulation serves as the measure of the initial node's spreading influence. Because the SIR model is stochastic, the simulation is repeated multiple times and the average outcome is taken as the final spreading value. Determining an appropriate infection probability is crucial: overly large values lead to near-uniform outbreaks, while very small values suppress the spreading process. Therefore, the threshold infection probability is computed for each network using $\beta_{th} = \frac{<k>}{<k^2>-<k>}$, ensuring meaningful and comparable diffusion dynamics across all datasets

**4-2- Dataset**

The dataset used in this study consists of nine real-world complex networks originating from diverse domains, including social interaction networks, transportation systems, technological infrastructures, and communication networks [26,27]. These networks vary significantly in size, density, and structural properties, offering a broad testbed for evaluating the performance of the proposed method. For each network, fundamental topological characteristics such as the number of nodes n, the number of edges m, the average degree ⟨k⟩, the average shortest-path distance ⟨d⟩, the clustering coefficient C, the degree assortativity r, and the epidemic threshold βth are reported in Table 2. These features enable robust and fair comparisons across networks with heterogeneous structures.

**4-3- Accuracy**

To assess how accurately a ranking method reflects the true spreading potential of nodes, its ranking list is compared against the ranking obtained from SIR simulations. For this purpose, the Kendall's Tau-b correlation coefficient is employed, as it quantifies the degree of agreement between two ordered lists [28]. The measure examines every pair of nodes x and y and determines whether their relative order is consistent or inconsistent across the two rankings. Unlike the basic Tau, the Tau-b variant also accounts for ties that may occur when several nodes receive identical ranks. The coefficient is computed using the following formulation:

$$Tau_b(R_1, R_2) = \frac{n_c - n_d}{\sqrt{(N - N_{t1})(N - N_{t2})}} \quad (9)$$

$$N = \frac{n(n-1)}{2}$$

$$N_{t1} = \sum_i \frac{t_i(t_i - 1)}{2}, N_{t2} = \sum_u \frac{t_u(t_u - 1)}{2}$$

Here, $n_c$ and $n_d$ denote the numbers of concordant and discordant node pairs, respectively. $N$ represents the total number of possible node pairs in a network of size $n$. The terms $t_i$ and $t_u$ are the counts of ties at rank iii in the first ranking $R_1$ and at rank u in the second ranking $R_2$. The value of $Tau_b$ ranges –1 to 1, where 1 indicates perfect agreement and –1 signifies completely reversed rankings. Higher Tau-b values indicate that the evaluated method produces rankings more consistent with the SIR-based ground truth.

Table 1- Dataset information

| Network | n | m | <k> | <d> | C | r | $\beta_{th}$ |
|---|---|---|---|---|---|---|---|
| Dolphins | 62 | 159 | 5.1 | 3.357 | 0.259 | -0.044 | 0.17 |
| Polbooks | 105 | 441 | 8.4 | 3.079 | 0.488 | -0.128 | 0.09 |
| USAir97 | 332 | 2126 | 12.8 | 2.738 | 0.625 | -0.208 | 0.02 |
| Netsci | 379 | 914 | 4.8 | 6.042 | 0.741 | -0.082 | 0.14 |
| WikiVote | 889 | 2914 | 6.6 | 4.096 | 0.153 | -0.029 | 0.06 |
| Email | 1133 | 5451 | 9.6 | 3.606 | 0.220 | 0.078 | 0.06 |
| Hamster | 2426 | 16631 | 13.7 | 3.589 | 0.538 | 0.047 | 0.03 |
| Router | 5022 | 6258 | 2.5 | 6.449 | 0.033 | -0.138 | 0.08 |
| PGP | 10680 | 24340 | 4.6 | 7.463 | 0.266 | 0.24 | 0.06 |

### 4-4- Resolution

To measure the resolution capability of a ranking method—that is, its ability to distinguish nodes with different spreading strengths—the Monotonicity Index is applied [29,30]. This index quantifies the extent to which a method assigns unique ranks to nodes. Its value lies between 0 and 1, with higher values indicating finer discrimination among nodes. The metric is defined as:

$$Mon(X) = [1 - \frac{\sum_{x \in X} N_x (N_x - 1)}{n(n-1)}]^2 \tag{10}$$

In this expression, $X$ denotes the ranking produced by a given method, $N_x$ is the number of nodes that receive the same rank $x$, and $n$ is the total number of nodes in the network. A method with minimal ties yields a Monotonicity value close to 1, reflecting high resolution in its ranking structure.

## 5. Results

### 5-1- Accuracy

Fig. 2 illustrates the accuracy of the proposed method across nine real-world complex networks, using a fixed radius of 3 for the QGC computation. The horizontal axis represents the spreading probabilities evaluated around the epidemic threshold, while the vertical axis shows the Kendall tau correlation. As the plots reveal, the proposed approach achieves the highest accuracy in eight of the networks when using R=3. In the Router network, QGC achieves the best accuracy up to 1.5 times the threshold, and for larger values, only GSM surpasses it. Overall, these observations demonstrate the robustness and effectiveness of the QGC method with a fixed radius of R=3 in almost all datasets.

### 5-2- Resolution

Table 2 reports the resolution scores of the recent methods and QGC, measured using the Monotonicity Index across the nine networks using a fixed radius of 3. The proposed method achieves the highest resolution in all networks, indicating that QGC not only ranks influential nodes accurately but also provides strong discriminative power even with a fixed radius.

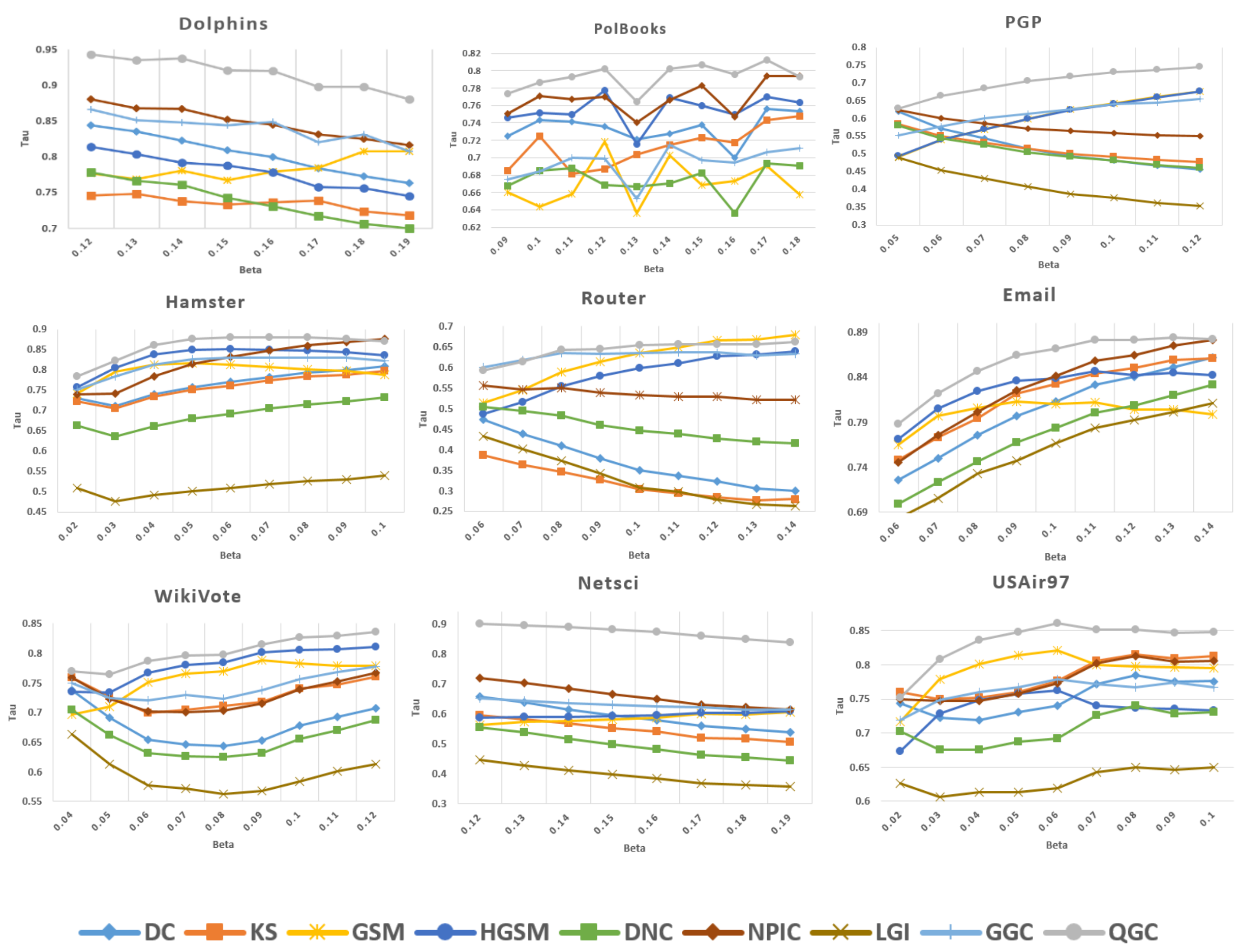

Fig 2- Kendall's Tau correlation coefficient in different spreading probabilities for nine methods in the dataset

Table 2- Monotonicity index of nine methods in the dataset

| Method | Dolphins | PolBooks | USAir97 | Netsci | WikiVote | Email | Hamster | Router | PGP |
|---|---|---|---|---|---|---|---|---|---|
| DC | 0.831 | 0.825 | 0.859 | 0.764 | 0.804 | 0.887 | 0.898 | 0.289 | 0.619 |
| KS | 0.377 | 0.495 | 0.811 | 0.642 | 0.727 | 0.809 | 0.871 | 0.069 | 0.481 |
| GSM | **0.998** | **1.000** | 0.994 | 0.995 | **1.000** | **1.000** | **0.986** | 0.996 | **1.000** |
| HGSM | **0.998** | **1.000** | **0.995** | 0.995 | **1.000** | **1.000** | **0.986** | **0.997** | **1.000** |
| DNC | 0.996 | **1.000** | **0.995** | 0.993 | 0.996 | 0.999 | 0.982 | 0.715 | 0.950 |
| QGC | **0.998** | **1.000** | **0.995** | **0.996** | **1.000** | **1.000** | **0.986** | **0.997** | **1.000** |
| NPIC | **0.998** | **1.000** | **0.995** | 0.995 | **1.000** | **1.000** | **0.986** | 0.996 | **1.000** |
| LGI | **0.998** | **1.000** | **0.995** | **0.996** | **1.000** | **1.000** | **0.986** | 0.995 | **0.998** |
| GGC | **0.998** | **1.000** | **0.995** | 0.995 | **1.000** | **1.000** | **0.986** | 0.996 | **1.000** |

## 5-3- Computational complexity

The proposed QGC method relies on two structural features: node degree and k-shell value, which can be computed in O(n) and O(m) time, respectively. Consequently, calculating HQC also requires O(m). In the second stage, the QGC score is computed by examining neighbors within a fixed three-hop radius for each node. Since HQC values have already been precomputed, this step requires $O(n < k >^3)$ where ⟨k⟩ is the average degree. Thus, the overall time complexity of the method is $O(n < k >^3)$ which is significantly lower than that of diffusion-based or deep-learning-based approaches.

### 5-4- Ablation analysis of the linear term in HQC

To investigate whether the quadratic term dominates the linear term $(k_i + ks_i)$ in heterogeneous networks, we conducted an ablation experiment by removing the linear component and retaining only the quadratic component . The results on nine benchmark networks are reported in Table 3. As shown in Table 3, eliminating the linear term does not lead to a consistent improvement across all networks. The modified version performs slightly better in some networks, such as Dolphins, PolBooks, Netsci, Email, Router, and PGP, whereas the original HQC achieves better results in USAir97 and WikiVote. In several cases, the differences are very small, indicating that the proposed formulation is stable. Overall, these results suggest that although the quadratic term is numerically dominant in some heterogeneous networks, the linear term still contributes to the generality and theoretical consistency of the proposed quasi-Laplacian formulation.

Table 3- Ablation study on the linear term in HQC.

| Network | HQC | Eliminate $(\boldsymbol{k_i} + \boldsymbol{ks_i})$ |
|---|---|---|
| Dolphins | 0.91676 | 0.91782 |
| PolBooks | 0.79571 | 0.79578 |
| USAir97 | 0.83382 | 0.83317 |
| Netsci | 0.87307 | 0.87328 |
| WikiVote | 0.80239 | 0.80219 |
| Email | 0.85788 | 0.85794 |
| Hamster | 0.85875 | 0.85876 |
| Router | 0.64148 | 0.64158 |
| PGP | 0.68300 | 0.69350 |

### 5-5-Rationale for the Fixed Gravity Radius R=3

This section explains the rationale for adopting a fixed gravity radius of R=3 throughout the manuscript. To evaluate the effect of radius size on the performance of the proposed method, the accuracy of QGC was examined for radius values from 1 to 4 across all benchmark networks, as illustrated in Fig. 3. For each network, the mean accuracy over the selected infection probability range was computed for different radius values, and the results are summarized in Table 4.

The results show that the effective interaction range lies within a narrow local scale, with R=2 and R=3 consistently performing better than both smaller and larger values. More specifically, R=3 achieves the highest mean accuracy in five of the nine networks, including USAir97, Netsci, WikiVote, Hamster, and Router, while R=2 gives the best result in three networks, namely Dolphins, PolBooks, and PGP. In the Email network, the performances of R=2 and R=3 are nearly identical. Importantly, even in the cases where R=2 performs slightly better, the difference between R=2 and R=3 is very small and does not affect the overall comparative conclusions.

These findings indicate that a very small radius such as R=1 is not sufficient to capture indirect structural effects beyond immediate neighbors, whereas a larger radius such as R=4 may introduce less relevant or redundant information from distant nodes. In contrast, R=3 provides a suitable balance between locality and structural coverage. It captures a wider neighborhood than R=2 while still preserving the local nature of influence propagation and maintaining low computational cost.

A meaningful relationship can also be observed between the effective radius and the average shortest-path length of the network. Networks with relatively small average shortest-path length, such as PolBooks and Dolphins, tend to achieve their best performance with R=2, whereas networks with more extended structures, such as Netsci and Router, benefit more from R=3. However, this trend is not absolute, which suggests that the effective radius is influenced not only by global path-length characteristics but also by local connectivity patterns and structural heterogeneity.

Based on these observations, R=3 is selected as the fixed radius for all QGC computations in this study. This choice is supported by three main reasons. First, R=3 yields the best performance in the majority of the tested networks. Second, in networks where R=2 is slightly superior, the difference is marginal. Third, using a fixed radius of R=3 avoids network-specific tuning and improves the reproducibility, simplicity, and general applicability of the proposed method. Therefore, R=3 is adopted as a robust and practical default setting for the QGC framework.

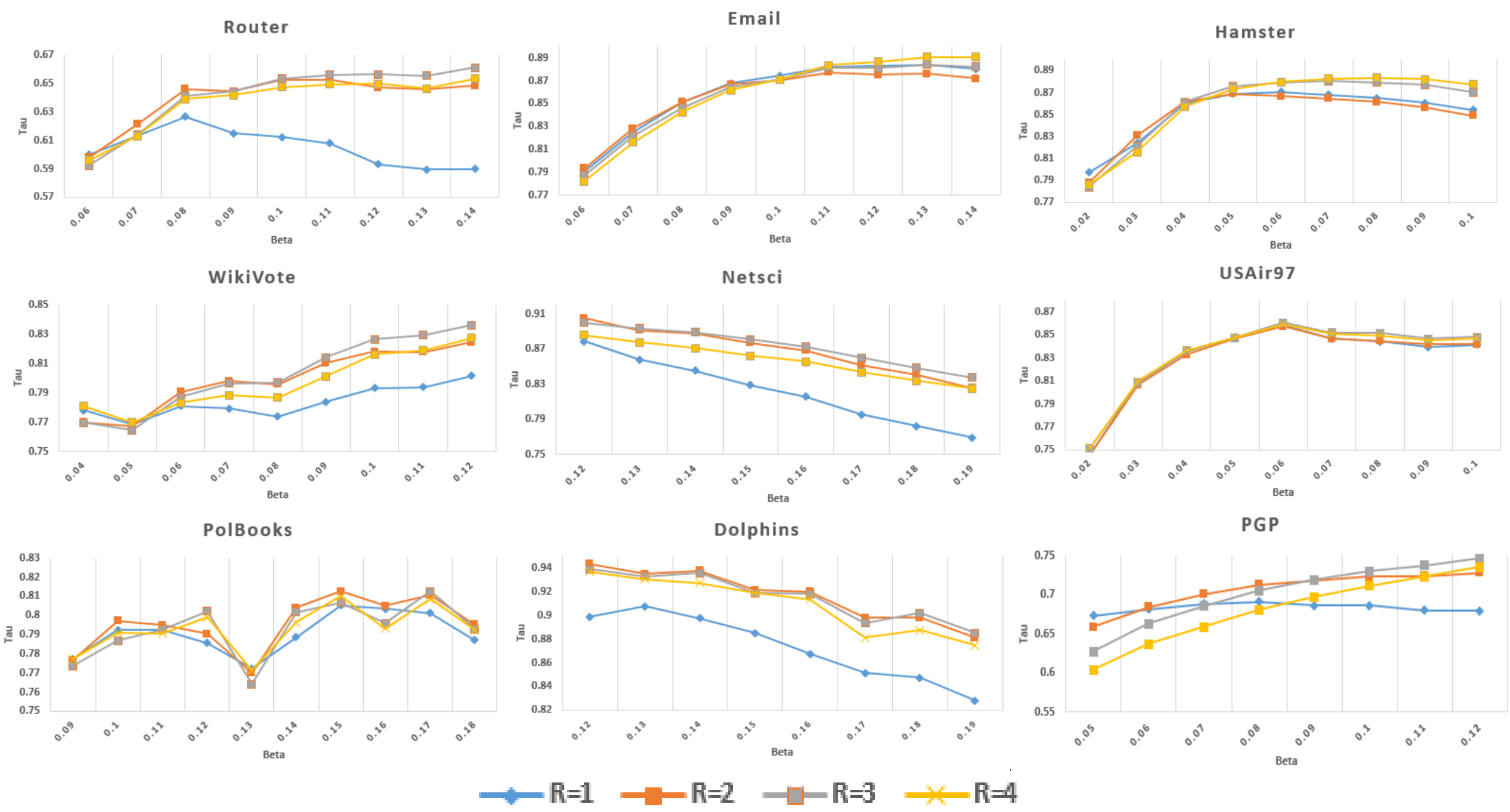

Fig 3- Sensitivity analysis of the proposed QGC method under different radius values.

Table 4. Mean accuracy of the proposed QGC method for different radius values across the benchmark networks.

| Network | R=1 | R=2 | R=3 | R=4 |
|---|---|---|---|---|
| Dolphins | 0.873 | **0.917** | 0.916 | 0.909 |
| PolBooks | 0.791 | **0.796** | 0.793 | 0.793 |
| USAir97 | 0.830 | 0.830 | **0.834** | 0.833 |
| Netsci | 0.822 | 0.869 | **0.873** | 0.857 |
| WikiVote | 0.784 | 0.799 | **0.802** | 0.797 |
| Email | 0.857 | 0.857 | **0.858** | 0.857 |
| Hamster | 0.852 | 0.850 | **0.859** | 0.859 |
| Router | 0.605 | 0.639 | **0.641** | 0.637 |
| PGP | 0.681 | **0.694** | 0.683 | 0.661 |

## 6. Conclusion and future work

This study introduced a hybrid approach for identifying influential nodes in complex networks by combining a quasi-Laplacian structural measure with a localized gravity model. The proposed HQC index integrates node degree, k-shell value, and the contribution of immediate neighbors, thereby capturing both local structural strength and hierarchical position. These HQC values are then used as network masses in the QGC framework, where distance-based weighting is applied within a fixed gravity radius of 3 to evaluate the short-range influence of nodes.

The experimental results on benchmark networks demonstrated that the proposed QGC method achieves strong performance in terms of both accuracy and resolution. The method is capable of identifying not only highly connected nodes but also structurally strategic nodes that play important roles in diffusion processes. In addition, by restricting the interaction range to a fixed local radius, the proposed method maintains relatively low computational complexity compared with simulation-based approaches, while preserving interpretability and avoiding network-specific parameter tuning.

Despite these advantages, the proposed method has several limitations. First, the HQC index relies on k-shell decomposition, which requires global knowledge of the network and repeated non-local updates during shell extraction. As a result, the method may be less suitable than purely local approaches for very large-scale or dynamically changing networks, and its effectiveness may also be reduced in networks with weak or ambiguous core-periphery organization. Second, although a fixed gravity radius of 3 provides a robust and reproducible setting across the tested networks, it may not be universally optimal for all network structures, especially for networks with very large diameter or highly heterogeneous connectivity patterns. As shown in the radius analysis, a few compact networks achieve slightly better performance with radius 2, although the difference remains small.

Several concrete directions may extend this work. First, a local approximation of k-shell could be developed to reduce the dependence on global network information and improve applicability to large or dynamic networks. Second, an adaptive radius selection mechanism based on structural properties such as average shortest-path length, local density, or clustering characteristics could be designed without relying on diffusion ground truth. Third, the framework can be generalized to weighted networks by replacing the degree with node strength and by incorporating weighted variants of shell-based structural measures. Fourth, a parallel implementation of the method for very large networks, for example through distributed or MapReduce-style computation, could further improve scalability. Finally, integrating the proposed framework with data-driven models may provide another promising direction, especially when empirical diffusion observations are available.

## 7. List of symbols

$k_i$: Degree of node i

$ks_i$: K-Shell of node i

$d_{ij}$: Shortest distance between node I and node j

$LC_j$: Local clustering coefficient of node j

$N_i$: Direct neighbors of node i

$n$: Number of nodes in the network

$m$: Number of edges in the network

$<k>$: Average Degree in the network

## References

[1] Boccaletti, S., Latora, V., Moreno, Y., Chavez, M., & Hwang, D. U. (2006). Complex networks: Structure and dynamics. Physics reports, 424(4-5), 175-308.

[2] Esfandiari, S. (2025, May). Enhancing Data Quality by Identifying Influential Nodes: Integrating Complementary Features with CoCoFISo. In Companion Proceedings of the ACM on Web Conference 2025 (pp. 2116-2119).

[3] Esfandiari, S., & Fakhrahmad, S. M. (2025). The collaborative role of K-Shell and PageRank for identifying influential nodes in complex networks. Physica A: Statistical Mechanics and its Applications, 658, 130256.

[4] Esfandiari, S., & Fakhrahmad, M. (2024, May). Predicting node influence in complex networks by the k-shell entropy and degree centrality. In Companion Proceedings of the ACM Web Conference 2024 (pp. 629-632).

[5] Esfandiari, S., & Moosavi, M. R. (2025). Identifying influential nodes in complex networks through the k-shell index and neighborhood information. Journal of Computational Science, 84, 102473.
[6] Esfandiari, S., & Fakhrahmad, S. M. (2024, April). Identifying influential users in social networks with an extended hybrid global structure model. In 2024 10th International Conference on Web Research (ICWR) (pp. 1-5). IEEE.
[7] Esfandiari, S., & Fakhrahmad, S. M. (2025). Identifying influential nodes in complex networks by adjusted feature contributions and neighborhood impact: S. Esfandiari, SM Fakhrahmad. The Journal of Supercomputing, 81(3), 503.
[8] Esfandiari, S. (2025, May). A Multi-Attribute Model to Identify Influential Nodes in Complex Networks: Addressing the K-Shell Limitations. In Companion Proceedings of the ACM on Web Conference 2025 (pp. 3092-3095).
[9] Esfandiari, S., & Fakhrahmad, S. M. (2024, February). Mining influential spreaders in complex networks by an effective combination of the degree and K-shell. In 2024 20th CSI International Symposium on Artificial Intelligence and Signal Processing (AISP) (pp. 1-6). IEEE.
[10] Esfandiari, S., & Fakhrahmad, S. M. (2026, April). Uncertainty-Aware Fuzzy Centrality Measures for Influential Node Identification: A Structural Modeling Approach Toward E-Commerce Applications. In 2026 12th International Conference on Web Research (ICWR) (pp. 173-178). IEEE.
[11] Sun, G. Q., He, R., Hou, L. F., Luo, X., Gao, S., Chang, L., ... & Zhang, Z. K. (2025). Optimal control of spatial diseases spreading in networked reaction–diffusion systems. Physics Reports, 1111, 1-64.
[12] Huang, D. W., Wu, W., Bi, J., Li, J., Gan, C., & Zhou, W. (2025). Timeliness-aware rumor sources identification in community-structured dynamic online social networks. Information Sciences, 689, 121508.
[13] Kaur, D., Kushwah, S., & Kumar, S. (2025). Viral marketing: a systematic literature review and future research agenda. Marketing Intelligence & Planning.
[14] Ullah, A., & Meng, Y. (2025). Finding influential nodes via graph embedding and hybrid centrality in complex networks. Chaos, Solitons & Fractals, 194, 116151.
[15] Hansen, D. L., Shneiderman, B., Smith, M. A., & Himelboim, I. (2020). Social network analysis: Measuring, mapping, and modeling collections of connections. Analyzing social media networks with NodeXL, 31-51.
[16] Kitsak, M., Gallos, L. K., Havlin, S., Liljeros, F., Muchnik, L., Stanley, H. E., & Makse, H. A. (2010). Identification of influential spreaders in complex networks. Nature physics, 6(11), 888-893.
[17] Barthelemy, M. (2004). Betweenness centrality in large complex networks. The European physical journal B, 38(2), 163-168.
[18] Ullah, A., Wang, B., Sheng, J., Long, J., Khan, N., & Sun, Z. (2021). Identification of nodes influence based on global structure model in complex networks. Scientific Reports, 11(1), 6173.
[19] Mukhtar, M. F., Abal Abas, Z., Baharuddin, A. S., Norizan, M. N., Fakhruddin, W. F. W. W., Minato, W., ... & Anuar, S. H. H. (2023). Integrating local and global information to identify influential nodes in complex networks. Scientific Reports, 13(1), 11411.
[20] Qiu, L., Zhang, J., & Tian, X. (2021). Ranking influential nodes in complex networks based on local and global structures. Applied intelligence, 51(7), 4394-4407.
[21] Li, H., Shang, Q., & Deng, Y. (2021). A generalized gravity model for influential spreaders identification in complex networks. Chaos, Solitons & Fractals, 143, 110456.
[22] Zhao, N., Yang, S., Wang, H., Zhou, X., Luo, T., & Wang, J. (2024). A novel method to identify key nodes in complex networks based on degree and neighborhood information. Applied Sciences, 14(2), 521.
[23] Ullah, A., Sheng, J., Wang, B., Din, S. U., & Khan, N. (2024). Leveraging neighborhood and path information for influential spreaders recognition in complex networks. Journal of Intelligent Information Systems, 62(2), 377-401.
[24] Patel, A., & Singh, B. (2025). Graph convolutional network for structural equivalent key nodes identification in complex networks. Chaos, Solitons & Fractals, 196, 116376.

[25] Weiss, H. H. (2013). The SIR model and the foundations of public health. Materials matematics, 0001-17.
[26] Rossi, R., & Ahmed, N. (2015, March). The network data repository with interactive graph analytics and visualization. In Proceedings of the AAAI conference on artificial intelligence (Vol. 29, No. 1).
[27] Jure, L. (2014). Snap datasets: Stanford large network dataset collection. Retrieved December 2021 from http://snap. stanford. edu/data.
[28] Dehling, H., Vogel, D., Wendler, M., & Wied, D. (2017). Testing for changes in Kendall's tau. Econometric Theory, 33(6), 1352-1386.
[29] Quesada, A. (2009). Monotonicity and the Hirsch index. Journal of Informetrics, 3(2), 158-160.
[30] Liberman, V., & Tversky, A. (1993). On the evaluation of probability judgments: Calibration, resolution, and monotonicity. Psychological Bulletin, 114(1), 162.